A SUPERNOVA AT 50 PC: EFFECTS ON THE EARTH'S ATMOSPHERE AND BIOTA

A.L Melott[1], B. C. Thomas[2], M. Kachelrieß[3], D.V. Semikoz[4,5], and A. C. Overholt[6]

1. Department of Physics and Astronomy, University of Kansas, Lawrence, Kansas 66045 USA; melott@ku.edu

2. Department of Physics and Astronomy, Washburn University, Topeka, Kansas 66621 USA

3. Institutt for fysikk, NTNU, Trondheim, Norway

4. APC, Universite Paris Diderot, CNRS/IN2P3, CEA/IRFU, Observatoire de Paris, Sorbonne Paris Cite, 119 75205 Paris, France

5. National Research Nuclear University "MEPHI" (Moscow Engineering Physics Institute), Kashirskoe highway 31, M4, 115409, Russia

6. Department of Science and Mathematics, MidAmerica Nazarene University, Olathe, Kansas, 66062 USA

ABSTRACT

Recent $^{60}$Fe results have suggested that the estimated distances of supernovae in the last few million years should be reduced from ~100 pc to ~50 pc. Two events or series of events are suggested, one about 2.7 million years to 1.7 million years ago, and another may at 6.5 to 8.7 million years ago. We ask what effects such supernovae are expected to have on the terrestrial atmosphere and biota. Assuming that the Local Bubble was formed before the event being considered, and that the supernova and the Earth were both inside a weak, disordered magnetic field at that time, TeV-PeV cosmic rays at Earth will increase by a factor of a few hundred. Tropospheric ionization will increase proportionately, and the overall muon radiation load on terrestrial organisms will increase by a factor of ~150. All return to pre-burst levels within 10kyr. In the case of an ordered magnetic field, effects depend strongly on the field orientation. The upper bound in this case is with a largely coherent field aligned along the line of sight to the supernova, in which case TeV-PeV cosmic ray flux increases are ~$10^4$; in the case of a transverse field they are below current levels. We suggest a substantial increase in the extended effects of supernovae on Earth and in the "lethal distance" estimate; more work is needed. This paper is an explicit followup to Thomas et al. (2016). We also here provide more detail on the computational procedures used in both works.

INTRODUCTION

We have used the expanded knowledge of Pliocene and Pleistocene epoch supernova irradiation available from observations and modeling published in 2016. In Thomas et al. (2016), we considered the effects of a supernova at about 100 pc, based on information from a series of new detections of $^{60}$Fe, particularly those in Wallner et al. (2016). Knie et al. (1999, 2004) had provided initial detections. Knie et al. (2004) initially suggested a distance of 40 pc, unlike most later researchers. Modeling and further detections showed consistency with an event beginning around 2.6 Ma (Ma = "Myr ago") (Fimiani et al. 2016; Breitschwerdt et al. 2016; Ludwig et al. 2016), probably preceded by other events.

Kachelrieß et al. (2015) showed that characteristics of ambient cosmic rays (CRs) can be explained by a supernova within a few hundreds of pc about 2 Ma. Such a source also explains characteristics of the CR dipole anisotropy (Erlykin & Wolfendale 2006; Savchenko et al. 2015). Based on Fry et al. (2015), we chose IIP as the best fit type.

Other results followed. Breitschwerdt et al. (2016), Wallner et al. (2016), Fimiani et al. (2016), Ludwig et al. (2016), and Binns et al. (2016) all provided strong confirmation of the general picture and new information. However, some recent assessments (Mamajek, 2016; a considerably refined modeling of transport and deposition by Fry et al. 2016) place the events at a more probable distance of ~50 pc, probably originating within the Tuc-Hor stellar group, which may have been at the appropriate distance at the appropriate time. We regard this factor of two difference in distance as completely reasonable, given the uncertainty in supernova yield of $^{60}$Fe and its transport, deposition, and uptake into sediments and magnetotactic bacteria. The purpose of this paper is to (1) explore the consequences of a nearer event, particularly for the important and nontrivial cosmic ray propagation, and (2) more fully describe our computational procedures, not previously possible within the restrictions of *Letters*.

There were no major mass extinctions in the last 10 Myr, but there is an elevated extinction in this period, formally classified as a mass extinction by Bambach (2006), but characterized as a pulse/turnover by Vrba (1992). There is some extinction connected with the generally cooling climate, which resulted in a transition to the Pleistocene period beginning about 2.6 Ma, with subsequent frequent glaciations. The situation is complicated by the timing of the closing of the Isthmus of Panama, which contributed to the extinction event by the change in ocean currents (O'Dea et al 2016; Stanley & Campbell 1981), and the interchange of fauna when North and South America joined around this time (Vrba 1992). Africa is a cleaner laboratory for the supernova effects, due to absence of the complicating geographic change. Vrba (1992) classified changes in vegetation cover as an extremely important variable in determining the species turnover on land. Evidence suggests a major shift in Africa from forest to semi-arid grasslands and a colder environment.

We can look for a possible hint of change in lightning rates and/or climate change brought on as a side effect of incident radiation (e.g Erlykin & Wolfendale 2010). There have been controversial suggestions of a strong link between CRs, clouds and climate (Svensmark 2015). However, as these links are not really understood (Mironova et al. 2015), we will confine ourselves to looking at (a) radiation on the ground and the upper ocean and (b) ionization of the atmosphere. We will provide ionization data, which can later be used for questions of lightning rates, as well as cloud formation and its possible effect on climate change.

In this paper we will treat two cases, A and B (see below) which are based on an event at 50 pc modeled as in Kachelriess et al. (2015) and Thomas et al. (2016). In all cases we are interested in the most recent event ~2.6 Ma. In case A we assume an intervening galactic magnetic field, with a regular plus turbulent component, and the regular component lying along the line of sight to the supernova. Moreover, we assume in this case that the Earth and the SN are connected by a magnetic field line. In this way, we maximise the CR flux and obtain an upper bound on the possible impact of a SN at 50 pc distance on the life on Earth. This case has no analog in Thomas et al. (2016). In case B, we assume that the supernova and the Earth both lie in the Local Bubble already excavated by previous supernova, with the regular component of the magnetic field swept away, and the turbulent field strongly reduced. Thus, only case B has parameters the same as Thomas et al. (2016), but the distance is halved.

Atmospheric effects by CRs are treated as in Melott et al. (2016) using tables from Atri et al. (2010); air shower modeling to derive muon and neutron fluxes at ground level in Atri and Melott (2011) and Overholt et al. (2013, 2015). We describe these in more detail below.

## 2. EFFECTS OF PHOTONS

### 2.1 GAMMAS AND X-RAYS

There are no detections of prompt gamma ray photons from Type II supernovae. A type IIP is thought to be an explosion inside an extended envelope, which converts much of the energy to visible light. As in Thomas et al. (2016) we find that gamma ray and Xray fluxes are too small for substantial effects, in this case by at least two orders of magnitude. It seems that a type IIP supernova would have to be at about 5 pc distance in order to have significant effects from these sorts of radiations. (It may be worth noting that Gehrels et al. (2003) estimated a lethal distance of 8 pc, using only photons for the estimate.) An event at 5 pc would be likely only at Gy intervals (Melott and Thomas 2011).

## 2.2 UV and VISIBLE LIGHT

Spectra available for SN2013ej and SN2012aw extend to the hard UV. Scaled to 50 pc distance, the irradiance would be approximately $4 \times 10^{-3}$ W m$^{-2}$ at the top of the atmosphere. This is several orders of magnitude smaller than the Solar UV irradiance at the top of the atmosphere and so is unlikely to have any significant impact on the atmosphere or significantly increase UV irradiance at Earth's surface.

There is a large body of evidence that enhanced illumination at night, especially in blue, is detrimental in a number of different ways to many different types of organisms; in 2015 the journal *Philosophical Transactions of the Royal Society B* devoted an entire issue to this topic (Gaston et al. 2015). Exposure to light during a normally dark period disrupts circadian rhythms, leading to loss of sleep, fatigue, changes in behavior, and other impacts detrimental to individuals and populations (Foster & Kreitzmann 2004; Longcore & Rich 2004; LeGates et al. 2013; Raap et al. 2015; Dominoni et al. 2016). Salmon (2003) and Witherington (1997) discuss disruptions in sea turtle reproduction caused by artificial lighting along beaches. Changes in tree frog foraging behavior have been observed at illuminations above $10^{-3}$ µW cm$^{-2}$ (Longcore & Rich 2004). Sky-glow light pollution has also been associated with disruption of reproduction in frogs and salamanders (Buchanan 1993; Jaeger & Hailman 1973; Rand et al. 1997), coral reef invertebrates (Sweeney et al. 2011), marsupial mammals (Robert et al. 2015), and birds (Kempenaers et al. 2010). Effects have even been noted in plants (Bennie et al. 2016).

Nighttime exposure to artificial lighting has been linked to cancers in animals, including rats (Vinogradova et al. 2009) and humans, including breast cancer (Stevens 2009; Schernhammer et al. 2001) and prostate cancer (Kloog et al. 2009). Evidence also supports a link between night-time illumination and suppression of immune systems (Arjona et al. 2009; Bedrosian et al. 2011).

Many of the physiological changes associated with nighttime lighting appear to be connected to melatonin production and includes insects (Foster et al. 2004; Haim & Zubidat 2015; Jones et al. 2015). Abundant evidence implicates light with wavelengths between 400 nm and 500 nm in these impacts, and others, both at night and otherwise (Brainard et al. 1984; Lockley et al. 2003; Vandewalle et al. 2007; Zaidi et al. 2007; Wood et al. 2013; Brüning et al. 2016; Marshall 2016).

Brainard et al. (1984) found that irradiance of 0.186 µW cm$^{-2}$ of "cool white" fluorescent light caused a decrease in pineal melatonin production in the Syrian hamster (*Mesocricetus auratus*), and that natural full moonlight irradiance as low as 0.045 µW cm$^{-2}$ caused pineal melatonin suppression in some animals, but not all. Lockley at al. (2003) found that 460 nm light at 12.1 µW cm$^{-2}$ irradiance caused a suppression of melatonin in humans, while a similar irradiance at 555 nm did not have a similar effect. Wood et al. (2012) found significant melatonin suppression in humans with 57-59 µW cm$^{-2}$ of light between 450 and 475 nm.

We examined available spectra of several Type IIP supernovae, including 1999em (Hamuy et al. 2001), 2005cs (Bufano et al. 2009), SN2012aw (Bayless et al. 2013), 2013ej (Valenti et al. 2014); spectral data were downloaded from the Weizmann Interactive Supernova data REPository (http://wiserep.weizmann.ac.il/). After scaling each spectrum to represent a supernova at 50 pc, we find that the irradiance between 400 and 500 nm is roughly between 0.12 and 0.24 µW cm$^{-2}$. This is comparable to the lower irradiance values shown to have some effect in animals, but significantly smaller than the values used in human melatonin suppression studies.

To summarize: Some physiological effects of blue light in the night sky would be significant for a few weeks. Otherwise visible and UV photons are of negligible importance.

## 3. EFFECTS OF CRs

### 3.1 SUMMARY OF PROPAGATION

The propagation of CRs in the Galaxy is determined by the regular and the turbulent component of the Galactic magnetic field (GMF) Berezinskii et al. (1990) is an excellent introduction and reference to the physics of CR propagation. The turbulent part of the magnetic field can be characterized by the power-spectrum P(k) and the correlation $l_c$ of its fluctuations with wavenumber k. Assuming a power-law $P(k) \sim k^{-\alpha}$ for the spectrum, the maximal length $L_{max}$ of the fluctuations and the correlation length $l_c$ are connected by $l_c = (\alpha - 1) L_{max}/(2\alpha)$. If moreover the turbulent field is isotropic and the regular field can be neglected, then the propagation of CRs can be described by an energy dependent scalar function, the diffusion coefficient D(E): Diffusion becomes isotropic if CRs propagate on distances $l \gg L_{max}$, while the energy dependence of the diffusion coefficient is determined by the slope α as $D(E) \sim E^{2-\alpha}$ for energies $E \ll E_{cr}$. Here, the critical energy is $E_{cr}$ defined by $R_L(E_{cr}) = l_c$ with $R_L$ as the Larmor radius of the charged CR. Thus the condition $E \ll E_{CR}$ ensures large-angle scattering, while the requirement $l \gg L_{max}$ guarantees that features of anisotropic diffusion are washed out.

Several detailed models for the regular component of the GMF exist and two of the most recent and detailed ones are the Jansson-Farrar (2012a,b) and the Pshirkov et al. (2011, 2013) models. Giacinti et al. (2015) compared both models and found that they lead qualitatively to the same predictions for CR propagation. An important constraint on CR propagation models comes from ratios of stable primaries and secondaries produced by CR interactions on gas in the Galactic disk. In particular, the Boron/Carbon ratio in the CRs has been recently measured by the AMS-02 experiment up to the rigidity 2.6 TV (Aguilar et al. 2016) and agrees nicely with the expectations for Kolmogoroff diffusion, α = 5/3. Using the Jansson-Farrar model, Giacinti et al. (2015) found that the B/C ratio can be reproduced choosing as the maximal length of the fluctuations $L_{max}$ = 25 pc and α = 5/3, if the turbulent field is rescaled to one tenth of the value proposed by Jansson-Farrar (2012a,b). For this choice of parameters, CR propagation in the Jansson-Farrar model reproduces a large set of local CR measurements and we will use this choice as our case A. Note also that while $l/L_{max} = 2$ only marginally satisfies the condition $l \gg L_{max}$, the dependence of CR propagation on the concrete realization of the turbulent field is negligible because of the dominating regular field.

In case A, we calculate the trajectories of individual CRs (emitted isotropically by the SN) using the code described and tested in Giacinti et al. (2012) which uses nested grids. For the calculation of the local CR flux, we have divided the nearby Galaxy into cells and saved the length of CR trajectories per cell. Since we are interested here in a nearby SN at the distance of 50 pc, we increased the number of cells compared to Kachelriess et al. (2015). We divided the Galactic plane into a non-uniform grid with cell size 20 pc X 20 pc at the position of the SN, which gradually increases to 100 pc X 100 pc at distances more than 500 pc. The vertical height of these cells was chosen as 20 pc. This allowed us to calculate simultaneously the flux at various places with differing perpendicular distance to the regular magnetic field line which goes through the SN. The maximal flux at Earth is obtained if a magnetic field line connects the SN and the Earth. In contrast to Thomas et al. (2016), we study now (in Case A) the case where a SN has the maximal impact on the Earth and place therefore the Earth on the GMF line going through the SN.

At present, the Earth is located in the Local Bubble, a region with diameter 100—200 pc characterized by a local under-density of matter. This bubble was excavated by stellar winds and SNe explosions. It is natural to assume that these plasma flows also partly expelled the regular and turbulent magnetic field. However, neither simulations for the creation of the Local Bubble which start from a realistic GMF model nor measurements of the magnetic fields inside the Local Bubble exist. One may speculate that the reduction of the turbulent field found by Giacinti et al. (2015) is partly a local effect caused by the Local Bubble.

Alternatively, one may assume that the SN explosion creating the Local Bubble had a more drastic effect on the local GMF structure, expelling the regular field almost completely and reducing the strength of the turbulent field. As case B, we assume a purely turbulent magnetic fields with strength B = 0.1 µG, as suggested e.g. by Avillez and Breitschwerdt (2005). Since the regular field vanishes, CRs propagate isotropically and we can apply the diffusion approximation with $D(E) = D_0 (E/E_0)^{1/3}$, $D_0 = 2 \times 10^{28}$ cm$^2$/s and $E_0 = 10$ GeV. The CR density n from a bursting source is then given by

$$n(E,r,t) = Q(E)/(\pi^{3/2} r_{diff}^3) \exp(-r^2/r_{diff}^2), \qquad (1)$$

where Q(E) is the source spectrum and $r_{diff}$ the effective diffusion distance, which is given by $r_{diff}^2 = 4 D t$. The CR intensity I then follows as $I = (c/4\pi) n$. In both cases, we assume that the SN injects instantaneously the CR energy $2.5 \times 10^{50}$ erg with source spectrum $Q(E) \sim E^{-2.2} \exp(-E/E_c)$ and cutoff energy $E_c = 1$ PeV.

The injection would be delayed until the remnant expands enough to release the CR; the release would not then be instantaneous but the approximation is adequate for our purposes. The total CR energy assumed is consistent with modern estimates of typical values for Type II supernovae (Higdon et al 1998; Kasen & Woosley 2009). It is effectively much larger than that of Gehrels et al. (2003), who approximated the CR flux 10 pc from a supernova by scaling up the normal Galactic Cosmic Ray (GCR) flux up by a factor of 100, partly motivated by the estimated energetics of SN1987a (e.g. Honda et al. 1989). We find from our explicit propagation computations given a typical supernova at 50 pc that this value is exceeded at that distance in cases A and B below. However, the CR flux could be arbitrarily low in the case of a coherent magnetic field transverse to the line of sight. So, the past magnetic field orientation is extremely important, given the primary importance of CR in determining the result.

The resulting CR intensities on Earth are shown in Fig.1a for case A and Fig.1b for case B respectively. In Figure 1 we plot flux*$E^2$ so that the area under the curve is proportional to the total energy between limits. Times are measured from the release of the CR, minus the photon travel time to the Earth. Case A here shows a very large increase, which is caused by the assumption of a field line connecting the Earth with the supernova, and cannot be compared with anything in Thomas et al. (2016). Case B here is consistent with a nearly fivefold increase over Case C in Thomas et al. (2016), not far off from an inverse-square scaling.

It can be seen that the major difference from the present day cosmic ray flux is the large amount of energy deposited by CRs in excess of a TeV. Due to the strong energy-dependence of the cosmic ray penetration in the atmosphere, this implies penetration and ionization at much greater atmospheric depths than is usual for cosmic rays or flux from solar events (e.g. Melott et al. 2016). Discussion of the effects follows, beginning with atmospheric computation procedures.

3.2 ATMOSPHERIC COMPUTATIONS

Atmospheric ionization can have important impacts on chemistry. Odd-nitrogen oxides ($NO_y$) are produced, leading to destruction of stratospheric ozone and subsequent rain-out of $HNO_3$ (Thomas et al. 2005). Ionization by SN CRs in this work was computed using tables from Atri et al. (2010), generated by CORSIKA. CORSIKA (COsmic Ray SImulations for KAscade) is a package combining high- and low-energy interaction models with a transport scheme for CRs in air. Tuned are the interaction models, to all kinds of accelerator and CR data The version used for atmospheric ionization was EPOS 1.61, UrQmd 1.3. This gives ionization rate (ions $cm^{-2}$ $s^{-1}$) as a function of altitude for different primary proton energies. The tables in Atri et al. (2010) give ionization for primaries with energy from 300 MeV to 1 PeV, for 46 altitude bins, from the ground to 90 km. In this work we used primary proton energies between 10 GeV and 1 PeV.

Atmospheric chemistry modeling was performed using the Goddard Space Flight Center (GSFC) 2D (latitude-altitude) chemistry and dynamics model. This model has been extensively tested for cases similar to the work presented here (Thomas et al. 2005, 2007, 2008; Ejzak et al. 2007). The model runs from the ground to 116 km in altitude, with approximately 2 km altitude bins, and from pole-to-pole in 18 bands of 10-degree latitude each. The model includes 65 chemical species, 37 transported species and "families" (e.g. $NO_y$), winds, small scale mixing, solar cycle variations, and heterogeneous processes (including surface chemistry on polar stratospheric clouds and sulfate aerosols). We use the model in a pre-industrial state, with anthropogenic compounds (such as CFCs) set to zero. It is important to note that some parameter values in the model (e.g. large scale winds and temperature fields) are prescribed and tied to modern-day values. While this prevents accurate simulation of the exact conditions at other geologic eras, our results are limited to comparisons between runs with and without external influence, rather than absolute quantities, which should remove most such uncertainties.

The ionization rate as a function of altitude for a particular CR case is generated by convolving the CR flux (essentially, number of incident CR protons at each energy) with the tables from Atri et al. (2010). The results are then translated into production of $NO_y$, using 1.25 $NO_y$ molecules per ion pair (Porter 1976) at all altitude levels. While not significant for this application, we also include production of $HO_x$ as described in Thomas et al. (2005). The chemical family production rate is read in to the GSFC atmospheric chemistry model and the model was then run for 10 years until steady-state conditions were reached, simulating a quasi-steady flux of CRs. We then examine the concentrations of various constituents, especially $O_3$, comparing a run with SN CR ionization input to a control run without that ionization source.

3.3 ATMOSPHERIC EFFECTS

In both cases, ionization in the lower troposphere exceeds that from normal galactic CRs (Figure 2). Both cases get a significant prompt cosmic ray flux within 100 years. In case A we find a factor of about $10^3$ increase in ionization, right down to the ground. In case B we find a factor of about 50 increase in ionization, also down to the surface. These increases as compared with results in Thomas et al. (2016) are consistent with and as expected from the cosmic ray results described near the end of Section 3.1. Increased ionization greater than or comparable to that caused by GCRs could be expected to persist for about 10 kyr in case B, or 1 Myr in case A. Of course, if there were a transverse ordered field, no such effects would be seen. Large ionization should cause an increase in lightning (Erlykin and Wofendate 2010; Mironiva et al. 2015). There have been controversial claims of a connection between atmospheric cloud formation and CRs (Svensmark 2015; Mironova et al. 2015). Since there are many open questions in this research, we cannot offer a conclusion, except to state our results and hope that they offer an opportunity for further research to clarify the question.

While a full analysis of the atmospheric chemistry changes induced by the SN CRs is beyond the scope of this work, we will discuss some results of the modeling described above. Ozone ($O_3$) is catalytically depleted by oxides of nitrogen (e.g. Thomas et al., 2005). The loss of ozone in the stratosphere has for decades been viewed as the main source of damage to the biota from astrophysical ionizing radiation events (e.g. Gehrels et al. 2003). In Figure 3 we show the globally averaged change in $O_3$ column density, comparing a run with SN CR ionization included to a run without that added ionization. Here we show both case A and B, at one time each. The times chosen (100 yr for case A, 300 yr for case B) correspond to those with the greatest increase in ionization in the stratosphere, where the greatest impact on $O_3$ column density occurs. So, the results in Figure 3 represent the maximum change in $O_3$ we would expect from the SN CRs.

When a new ionization flux is introduced, it takes the atmosphere 3-5 years to equilibriate at the new value (Ejzak et al. 2007). However, in our cases the changes in CR flux are generally much more gradual than that. In Figure 3 we show plots of the globally averaged change in $O_3$ column density, assuming the sudden onset of the peak CR flux value found in Figure 1 to a normal atmosphere. Thus the depletion reached asymptotically after a few years should represent the depletion level from a slowly varying CR flux. In case A (at 100 y) this is about 66% depletion; in case B (at 300 y) it is 25%; in the case of a transverse magnetic field it should be negligible. We note that our case B at 50 pc is roughly comparable to the values found by Gehrels et al. (2003) for a 10 pc distance; thus we have significantly revised upward the CR atmospheric effects from that publication. For reference, recent anthropogenic $O_3$ depletion is a few percent, globally averaged.

Reduction in $O_3$ column density is important for life on Earth's surface and in the oceans since $O_3$ is a strong absorber of solar UV. Past work modeling effects of nearby gamma-ray bursts found around 35% globally averaged depletion (Thomas et al. 2005), which was suggested to be "mass extinction level." More recent work (Thomas et al. 2015, Neale & Thomas 2016) has shown that this may be overly pessimistic and that the biological impact is more complex, depending heavily on particular types of organisms and impacts considered. Nonetheless, globally averaged depletions on the order 60% surely would have an impact.

While a full analysis of the biological impact is beyond the scope of this work, we can make a simple estimate of the increase in biologically active solar UV due to a reduction in $O_3$ column density. Madronich et al. (1998) developed a power law relationship between various biological effects and $O_3$ column density: $UV_{bio} \sim (O_3)^{-RAF}$, where RAF is a "radiation amplification factor," which varies depending on the specific effect considered. Skin damage (erythema) and skin cancer have RAF values around 1.5, while DNA damage has RAF around 2. We can estimate an enhancement in biologically damaging UV using $(O_3'/O_3)^{-RAF}$, where the prime indicates the $O_3$ column density in the SN case. For case A at 100 yr the ratio is about 0.32; with RAF values 1.5 and 2.0, the resulting increase in UV is about 5 to 10 times normal, respectively. For case B the increase is about 1.5 times.

While the resulting biological effects are complicated and the ecological impact of any such effects are uncertain, it seems likely that a 5-10 times increase in biologically active UV would have a major impact. It is usually assumed that a globally averaged decrease of 30-50% would have an impact of mass extinction level. Based on that assumption and modeling performed by work by Gehrels et al. (2003), the "lethal distance" for a SN has been estimated at about 10 pc. If this is an appropriate measure of extinction potential, then our results move this "lethal distance" to 50 pc or more, *but only for interstellar conditions similar to that in our case A.* For conditions similar to our case B, that distance would be somewhat less than 50 pc, but still larger than 10 pc.

There was not a major mass extinction at the end of the Pliocene. Why not? There are several possibilities: (1) There was an ordered component of the galactic magnetic field transverse to the line of sight to the supernova(e). This would reduce the CR flux, possibly bringing it down to a negligible enhancement in the extreme limit. We consider this unlikely, due to the energetics of the Local Bubble, but it is possible. (2) The distance estimate used here was wrong, All acknowledge considerable uncertainty in the distance estimates. (3) There was a disordered magnetic field as in case B. This is the most likely condition given the recent $^{60}$Fe data. In this case there would be a 50% increase in UVB, dangerous but not major mass-extinction levels. (4) Recent detailed studies of phytoplankton productivity in an environment simulating astrophysical ionizing radiation changes in atmospheric transmission have suggested that effects are not as strong as assumed in the past, for a variety of reasons (Neale and Thomas 2016). That work studied only primary productivity, but at the least it indicates that the situation is more complicated than previous work assumed, and that more study is needed before setting a "lethal distance" for supernovae. Our results support the need for more study of this question, since no major mass extinction is observed to coincide with the time of the supernovae. Moderate UV effects could have played a role in the moderate extinction observed then.

The nitrogen oxides produced by ionization in the atmosphere are removed over a time scale of months to years by rain/snow-out of HNO$_3$.  At steady state in our modeling the global annual average deposition above background is about 0.5 g m$^{-2}$ for case A at 100 yr, and 0.03 g m$^{-2}$ for case B at 300 yr.  These values are about 80 times and 5 times, respectively, that generated by lightning and other nonbiogenic sources in today's atmosphere (Schlesinger 1997).  Such an increase is unlikely to have a detrimental effect (Thomas and Honeyman 2008) and in fact would probably be beneficial by providing nitrate fertilizer to primary producers.  Thomas and Honeyman (2008) and Neuenswander and Melott (2015) found that deposition of order 0.09 g m$^{-2}$ following a GRB would likely have a small fertilizer effect on both terrestrial and aquatic life.  Therefore, we expect that the deposition found in both cases considered here is very likely to have such an impact.

3.4 RADIATION ON THE GROUND: COMPUTATIONS

The majority of particles produced in cosmic ray showers do not penetrate to ground level. These particles lose energy through ionizing the atmosphere and do not reach sea level in sufficient number to produce an increase in average radiation level. There are two species of particles which produce sufficient amounts of cosmogenic radiation at sea level to be easily detected: muons and neutrons.

Muons penetrate to sea level and even below sea level through ocean water and rock. Muons are primarily produced high in the atmosphere as hadron-hadron interactions of cosmic rays produce pions which quickly decay into muons. Muon fluxes at ground level were found by convolving the primary spectra with the table of Atri & Melott (2011). The tables of Atri & Melott (2011) were produced again using CORSIKA, EPOS 1.99, FLUKA 2008. To reduce computation time, the energy cut for the electromagnetic component was set at 300 MeV. This value should be adequate to produce all hadronic interactions from protons while greatly reducing computational load (Atri & Melott 2011). After convolution, these tables produce a muon spectrum at ground level which can then be used to find the radiation dose. Radiation doses due to muons have not been calculated for a large range of specific energies as is needed for this work, so approximations must be used. To find the dose from our target muon flux, the energy lost by high energy muons as they travel through water was used (Klimushin et al. 2001). This approximation assumes the energy lost by muons as they travel through water is similar to the energy lost by traveling through biologic matter. Muons behave similar to a heavy electron, and in many cases can be approximated as such. For this reason, we use a radiation-weighting factor of 1, which is the same as an electron. Some research suggests that this approximation is inadequate for doses at skin level, but is adequate for the dose experienced by deeper organs (Siiskonen 2008). Through these two approximations, we develop an approximate radiation dose for cosmogenic muons from a given primary spectrum. When the average background cosmic ray primary spectrum is used, this method agrees within statistical error with the average radiation background from muons.

Neutrons are predominantly created higher in the atmosphere; however, their lack of charge allows them to penetrate substantially lower before thermalization. Neutrons interact through collisions with nuclei, losing energy with each collision. Once reaching thermal speeds, neutrons remain in the region in which they reach this speed, often in the stratosphere. Some neutrons reach this speed near or at sea level, producing a radiation dose for organisms on the ground. Neutron fluxes at ground level were found by convolving the primary spectrum through the tables of Overholt et al. (2013). These tables were produced from CORSIKA simulations identical to those used to find the muons, with the additional step of using MCNP (Monte Carlo N-Particle Simulator) to simulate low energy neutron interactions (Overholt et al. 2013). Here MCNP was used to simulate all neutrons below 50 *MeV* with order of magnitude bins as to produce sufficient particle flux for minimized statistical error (Overholt et al. 2013). Radiation doses per neutron were found by taking order of magnitude averages of the ionizing radiation dose per neutron from Alberts et al. (2001). These averages were then multiplied by the neutron fluxes at ground level to find the total neutron radiation dose.

3.5 RADIATION ON THE GROUND: RESULTS

We have determined the flux of muons on the ground as a function of energy. The detailed spectrum is not extremely important as the dose for muons only varies slowly with energy. We are normally exposed to an average of 56 muons $(m^2 \text{ sr s})^{-1}$. In the same units, at 100 y, in case A, the total flux is ~4.5 X $10^4$; in case B it is ~3.5 X $10^3$. The energy spectrum of the muons incident at sea level at this time is shown in Figure 4.

In Table 1, we show the muon and neutron doses on the ground for cases A and B. We must compare the increase due to muons and/or neutrons to the total background radiation dose. The radiation dose is capable of producing damage to terrestrial biota; emphatically in case A. The radiation doses exceed the average worldwide radiation background dose from all sources (2.4 mSv $yr^{-1}$, UNSC 2008) for hundreds to thousands of years, and would be likely to cause increased cancer risk and mutations. The ICRP (International Commission on Radiological Protection) models the risk of getting cancer increases by 5.5% for every 1 Sv of exposure (ICRP 1991). By comparison, the ICRP finds that an annual radiation dose of 1 mSv will only result in a very small health risk after a lifetime of exposure at this level (ICRP 1991). The dose sustained on the Earth during the time period following a supernova would build up a dose of 1 Sv in a time period from about two years (in Case A), to about 30 years (in Case B). Muons are more penetrating than most terrestrial radiation sources, so the change would also be more important in large organisms. Increasing the cancer risk of short lived (< 1 year) organisms by ~5% and long lived organisms by a larger factor could have detectable effects on the ecosystem as a whole. Figure 5 shows the time evolution of the dose from muons and neutrons, with horizontal bars to indicate the present level. The effect of neutrons is much less severe than muons. However, their high LET radiation does have the opportunity to produce more double strand DNA breaks, which can be the cause of more congenital malformations (Overholt et al. 2015). The time for return to normal levels is significantly different for Case A (upper bound of shaded region) and Case B (lower bound of shaded region). We note that the muon dose in Case B is about five times larger than the dose in Case C in Thomas et al. (2016). Due to different assumed magnetic field configuration in our Case A here, Case B is the only one that can be directly compared with our previous work, and does correspond approximately to simple inverse-square expectations—in fact better than might be expected in this diffusive approximation.

## 3.5 DIRECT DEPOSITION AND COSMOGENIC ISOTOPES

We must consider the possibility of radioactivity on the ground. As discussed by Ellis et al. (1996), there are two primary sources.

Cosmogenic isotopes such as $^{10}$Be and $^{14}$C are generated by the actions of CRs in the atmosphere. Using the discussion of Ellis et al. (1996) and considering a source at ~50 pc, we would expect of order $4 \times 10^5$ atoms cm$^{-2}$ of radionuclides during the passage of CRs generated by events A or B.

Direct deposit of SN-generated isotopes from a ~10 M$_\odot$ event at 50 pc can be expected, according to Ellis et al. (1996) to deposit ~ a few $\times 10^6$ atoms cm$^{-2}$ assuming that the interstellar medium has been previously cleared by another event (case B); if not they estimate that the blast wave would stall at about 50 pc (case A). Amounts of various isotopes are tabulated in Ellis et al. (1996), Table 1. These amounts are too small to be significant for the environment, but the direct deposit amounts lie within range of detection (e.g. Wallner et al. 2016). The 2.6 Ma age of the most recent event indicated by $^{60}$Fe is too great to have any prospect of detection in ice cores.

## 4. SUMMARY OF EFFECTS AND DISCUSSION

Common to all cases we note a possibly significant increase in blue light for a few weeks, which could be damaging to the biota if the event were visible in the night sky.

This would not likely be at all detectable in the fossil record. Effects of direct UV, X-rays, and gamma rays from the IIP supernova or its remnant appear to be insignificant.

Cosmic rays, however, may be extremely important, but their propagation is modulated by magnetic fields. Our case A has a primarily ordered magnetic field along the line of sight to the supernova. A variant on this has a transverse ordered 1 µG magnetic field. In this case the CR transport is so suppressed that we will not discuss it further. Case B, favored by recent $^{60}$Fe detections and the physics of the Local Bubble, suggests a primarily disordered magnetic field ~0.1 µG.

Atmospheric ionization by CRs depletes stratospheric ozone in the widely studied manner, resulting in ~66% ionization in case A and ~25% in case B. A caveat on case A is that it is near the limit of validity of the computational atmospheric chemistry model. The case A ionization level would have been labeled mass extinction level in past work, but that has been called into question by recent experimental studies. The case B level would be somewhat dangerous, probably triggering some increase in cancer, but difficult to see in the fossil record (Natarajan et al. 2007).

Because of the presence of much higher kinetic energy CRs, a major difference from any other recently studied astrophysical ionization mechanism operative at Earth (commonly solar proton events and gamma-ray bursts), is that atmospheric ionization proceeds right down to the ground—up to a factor of $10^3$ in Case A and a factor of 50 in Case 5. These levels can persist for hundreds to thousands of years. Previous work in this area has not considered tropospheric ionization, concentrating on stratospheric ionization which ultimately depletes the ozone layer. Consequently, there has not been exploration of the consequences of such ionization. Effects should include increased lightning, a nitrogen fertilizer effect from nitrate rainout (Thomas & Honeyman 2008; Neuenswander & Melott 2015), and possibly cloud formation in the troposphere. Past work which considered the effect of cosmic ray fluctuations on climate (e.g. Mironova et al. 2015) did not usually consider a major increase in ionization right down to the bottom of the troposphere. This deserves more attention.

Muon irradiation on the ground will increase by a factor of about 2000 for about a hundred years in case A, and a factor 150 in case B. This will cause an order of magnitude increase in the average overall irradiation at sea level even in case B, and include penetration of up to a kilometer in the oceans. A modest increase in the mutation rate is possible, and possible measurable ill effects, particularly in large animals, as the muons are much more penetrating than other radiation backgrounds to which they are normally exposed. This could have caused an observed disparity between fossil and molecular dates (Melott 2017), as well as facilitated a more rapid evolutionary response to the habitat changes which are credited with pushing the major faunal turnover at the beginning of the Pleistocene.

Data: the cosmic ray spectra at various times as published in this paper, and in Thomas et al. (2016) are available for download from Washburn University at http://www.washburn.edu/faculty/bthomas/research_data/ and at NTNU at http://web.phys.ntnu.no/~mika/SN_spectra.html . For details see these sites and the README files within.

We thank Bruce Lieberman and two referees for extremely helpful comments. BCT, ALM, and ACO gratefully acknowledge the support of NASA Exobiology grant NNX14AK22G. This research was supported in part with computational resources at NTNU provided by NOTUR, http://www.sigma2.no. Computation time at Washburn University was provided by the High Performance Computing Environment (HiPACE); thanks to Steve Black for assistance.

REFERENCES

Aguilar et al., [AMS collaboration] 2016, Phys. Rev. Lett. 117, 231102.

Alberts, W. G., Bartlett, D. T., Chartier, J. L., Hirning, C. R., McDonald, J. C., Schraube, H., Schwartz, R. B., 2001, J. of the ICRU, 1(3).

Arjona, A. et al. 2012, Trends Immunol., 33, 607

Atri, D. & Melott, A.L. 2011, Rad. Phys. Chem. 81, 701.

Atri. D. et al. 2010, JCAP, 008, doi:10.1088/1475-7516/2010/05/008.

Avillez, M.A. & Breitschwerdt, D. 2005 A&A 436, 585. DOI: 10.1051/0004-6361:20042146

Bayless et al. 2013, ApJL, 764, L13.

Bedrosian, T.A. et al. 2011, Biol. Lett., 7, 468.

Berezinskii, V. S., Bulanov, S. V., Dogiel, V. A., Ginzburg, V. L., and Ptuskin, V. S. 1990, Astrophysics of Cosmic Rays. North-Holland, Amsterdam.

Amsterdame, J. et al. 2016, J. Ecology, doi: 10.1111/1365-2745.12551.

Binns, W. R. et al. 2016, Science 352, 677-680. DOI:10.1126/science.aad6004

Brainard, G.C. et al. 1984, J. Pineal Res., 1, 105

Breitschwerdt, D. et al. 2016, Nature 532, 73-76.

Brüning, A. et al. 2016, Science of the Total Environment, 543, 214

Buchanan, B.W. 1993, Anim. Behav., 45, 893

Bufano et al. 2009, ApJ, 700, 1456.

Chakraborti, S. et al., 2016, ApJ, in press. arXiv:1510.06025v2.

Dominoni, D.M. et al. 2016, Biol. Lett., 12, 20160015.

Dwarkadas, V.V. 2014, MNRAS 440, 1917.

Eichler, D., Kumar, R., & Pohl, M. 2013, ApJ 769, 138.

Ejzak, L.M. et al., 2007, ApJ 654, 373.

Ellis, J. et al. 1996, ApJ 470, 1227

Erlykin, D., & Wolfendale, A.W. 2006, Astropart. Phys. 25, 183.

Erlykin, D. & Wolfendale, A.W. 2010, Surv. Geophys. 31, 383.

Fimiani, L. et al. 2016, Phys. Rev. Lett. 116. DOI: 10.1103/PhysRevLett.116.151104

Foster, R.G. & Kreitzmann, L. 2004, Rhythms of life: the biological clocks that control the daily lives of every living thing (New Haven, CT: Yale U. Press)

Fry, B. J. et al. 2016 ApJ 828, 48.

Gaston, K.J. et al. 2015, Phil. Trans. R. Soc. B, 370, 20140133.

Gehrels, N. et al. 2003, ApJ 585, 1169.

Giacinti, G., Kachelrieß, M., Semikoz, D.V., Sigl, G. 2012b, JCAP 1207, 031.

Giacinti, G., Kachelrieß, M., Semikoz, D.V. 2015, Phys. Rev. D 91, 083009.

Haim, A.H. & Zubidat, A.E. 2015, Phil. Trans. R. Soc. B, 370, 20140121

Hamuy, M. et al. 2001, ApJ, 558, 615.

Higdon, J.C., Lingefelter, R.E., & Ramaty, R. 1998, ApJL 509, L33.

Honda, M., Sato, H., & Terasawa, T. 1989 Prog. Theor. Phys. 82, 315.

Huang, F. et al. 2015, ApJ, 807, 59.

ICRP 21(1-3), 1991.

Jaeger, R.G. & Hailman, J.P. 1973, Z. Tierpsychol., 33, 352

Jansson, R. & Farrar, G.R. 2012a, Astrophys. J. 757, 14.

Jansson, R. & Farrar, G.R. 2012b, Astrophys. J. 761, L11.

Jones, T.M. et al. 2015, Phil. Trans. R. Soc. B, 370, 20140122

Kachelrieß, M. et al. 2015, Phys. Rev. Lett 115, 181103.

Kasen, D., & Woosley, S.E. 2009, ApJ 703, 2205.

Kempenaers, B. et al. 2010, Curr. Biol., 20, 1735

Klimushin, S., Bugaev, E., & Sokalski, I., 2001, Proc. of the 27th Intl. CR Conf. 07-15 August, 2001. Hamburg, Germany, 1009.

Kloog, I. et al. 2009, Chronobiology Int., 26, 108

Knie, K. et al. 1999, Phys. Rev. Lett. 83, 18. Doi:10.1103/PhysRevLett.83.18

Knie, K. et al. 2004, Phys. Rev. Lett. 93, 171103. https://doi.org/10.1103/PhysRevLett.93.171103

Longcore, T. & Rich, C. 2004, Front. Ecol. Environ. 2, 191-198.

Ludwig, P., et al. 2016, PNAS 113, 9232. doi:10.1073/pnas.1601040113

Mamajek, E.E. 2016 Proceedings of IAU Symposium 314, 21.

Marshall, J. 2016, Eye, 30, 2

Melott, A.L. 2017 Astrobiology 17, 87. doi:10.1089/ast.2016.1527

Melott, A.L. & Bambach, R.K. 2013, ApJ 773, 6. doi:10.1088/0004-637X/773/1/6

Melott, A.L. & Thomas, B.C. 2009, Paleobiology 35, 311. doi:10.1666/0094-8373-35.3.311

Melott, A.L. & Thomas, B.C. 2011, Astrobiology 11, 343. doi:10.1089/ast.2010.0603.

Melott, A.L. et al. 2016, JGR Atmospheres 121. DOI: 10.1002/2015JD024064

Mironova, I.A. et al. 2015, Spa. Sci. Rev. 194, 1. DOI 10.1007/s11214-015-0185-4

Natarajan, A.L., et al. 2007, Transactions of the Kansas Academy of Science 110, 155.

Neale, P.J., & Thomas, B.C. 2016, Astrobiology 16, 245. DOI: 10.1089/ast.2015.1360

Neuenswander, B. & Melott, A. 2015, Adv. Sp. Res. 55, 2946. DOI: 10.1016/j.asr.2015.03.017

O'Dea, A. et al. 2016, Science Advances 2, e1600883. DOI: 10.1126/sciadv.1600883

Overholt, A.C. et al. 2013, JGR Space Physics 118, 3765. doi:10.1002/jgra.50377

Overholt, A.C. et al. 2015, JGR Space Physics 120. doi:10.1002/2014JA020681.

Porter, H.S., Jackman, C.H., & Green, A.E.S., 1976, J. Chem. Phys., 65, 154

Pshirkov, M.S., et al. 2011, ApJ 738, 192.

Pshirkov, M.S., Tinyakov, P.G., and Urban, F.R. 2013, MNRAS 436, 2326.

Raap, T. et al. 2015, Sci. Rep., 5, 13557

Rand, A.S. et al. 1997, Copeia, 1997, 447

Robert, K.A. et al. 2015, Proc. R. Soc. B, 282, 20151745

Salmon, M. 2003, Biologist, 50, 163

Savchenko, V., Kachelrieß, M. & Semikoz, D.V. 2015, Astrophys. J. 809, L23.

Schernhammer E.S., et al. 2001, J Natl Cancer Inst., 93, 1563

Schlesinger, W.H. 1997, Biogeochemistry (2nd ed.; San Diego: Academic)

Siiskonen, T., 2008, Radiation Prot. Dos., 128(2), 234-238.

Stanley, S.M. & Campbell, A.D. 1982 Nature 293, 457

Stevens, R.G. 2009, Int. J. Epidemiol., 38, 963

Svensmark, H. 2015, Europhys. News. 2, 26.

Sweeney, A.M. et al. 2011, J. Exp. Bio., 214, 770

Thomas, B. C. et al. 2005, ApJ 634, 509.

Thomas, B.C., Jackman, C.H., & Melott, A.L. 2007, Geophys. Res. Lett., 34, L06810

Thomas, B.C. & Honeyman, M.D. 2008, Astrobiology, 8, 731. DOI: 10.1089/ast.2007.0262

Thomas, B.C., Melott, A.L., Fields, B.D., & Anthony-Twarog, B.J. 2008, Astrobiology, 8, 9

Thomas, B.C., Melott, A.L., Arkenberg, K.R., & Snyder II, B.R. 2013, Geophys. Res. Lett., 40, 1

Thomas, B.C. et al. 2015, Astrobiology 15, 207. doi:10.1089/ast.2014.1224.

Thomas, B.C. et al. 2016, ApJL 826, L3.  http://dx.doi.org/10.3847/2041-8205/826/1/L3

UNSC on the Effects of Atomic Radiation 2008. New York: United Nations, 4. ISBN 978-92-1-142274-0

Usoskin, I.G. et al. 2011, Atmos. Chem. Phys., 11, 1979-2011, doi:10.5194/acp-11-1979-2011.

Valenti et al. 2014, MNRAS 438, L101

Vandewalle, G. et al. 2007, PLoS ONE, 2, e1247

Vinogradova, I.A. et al. 2009, Aging, 1, 855

Vrba, E.S. 1992, J. Mamm., 73,1-28

Wallner, A. et al. 2016, Nature 532, 69-72.

Witherington, B.E. 1997, in Behavioral approaches to conservation in the wild, ed. J.R. Clemmons & R. Buchholz (Cambridge, UK: Cambridge U. Press), 303

Wood, B. et al. 2013, Applied Ergonomics, 44, 237

Xing, Y. et al. 2016, arXiv:1603.00998 [astro-ph.HE]

Zaidi, F.H. et al. 2007, Current Biology, 17, 2122

| Time | Muon Dose Case A | Neutron Dose Case A | Muon Dose Case B | Neutron Dose Case B |
|---|---|---|---|---|
| Present | 1.95E-01 | 1.20E-03 | 1.95E-01 | 1.20E-03 |
| 100yr | 4.37E+02 | 4.26E-01 | 3.15E+01 | 1.66E-02 |
| 300yr | 2.83E+02 | 2.48E-01 | 2.11E+01 | 2.99E-02 |
| 1kyr | 1.54E+02 | 1.37E-01 | 8.36E+00 | 2.20E-02 |
| 3kyr | 7.74E+01 | 7.82E-02 | 2.61E+00 | 9.37E-03 |
| 10kyr | 3.75E+01 | 4.66E-02 | 5.87E-01 | 2.53E-03 |
| 30kyr | 1.70E+01 | 2.16E-02 | 1.33E-01 | 6.27E-04 |
| 100kyr | 6.52E+00 | 1.09E-02 | 2.42E-02 | 1.21E-04 |
| 300kyr | 1.76E+00 | 5.13E-03 | 4.91E-03 | 2.52E-05 |
| 1Myr | 3.40E-01 | 1.69E-03 | 8.33E-04 | 4.36E-06 |
| 3Myr | 6.74E-02 | 3.24E-04 | 1.63E-04 | 8.61E-07 |

Table 1
Annual ground level secondary radiation dose, in units of mSv, due to muons and neutrons, for our two SN cases at various different times (measured from the arrival time of the first cosmic rays released by the SN remnant).  For comparison, the present day average annual dose on the ground from muons is about 0.20 mSv; from neutrons it is about 0.17 mSv; from all sources it is about 2.4 mSv. In both cases the muon dose is significant for thousands of years.

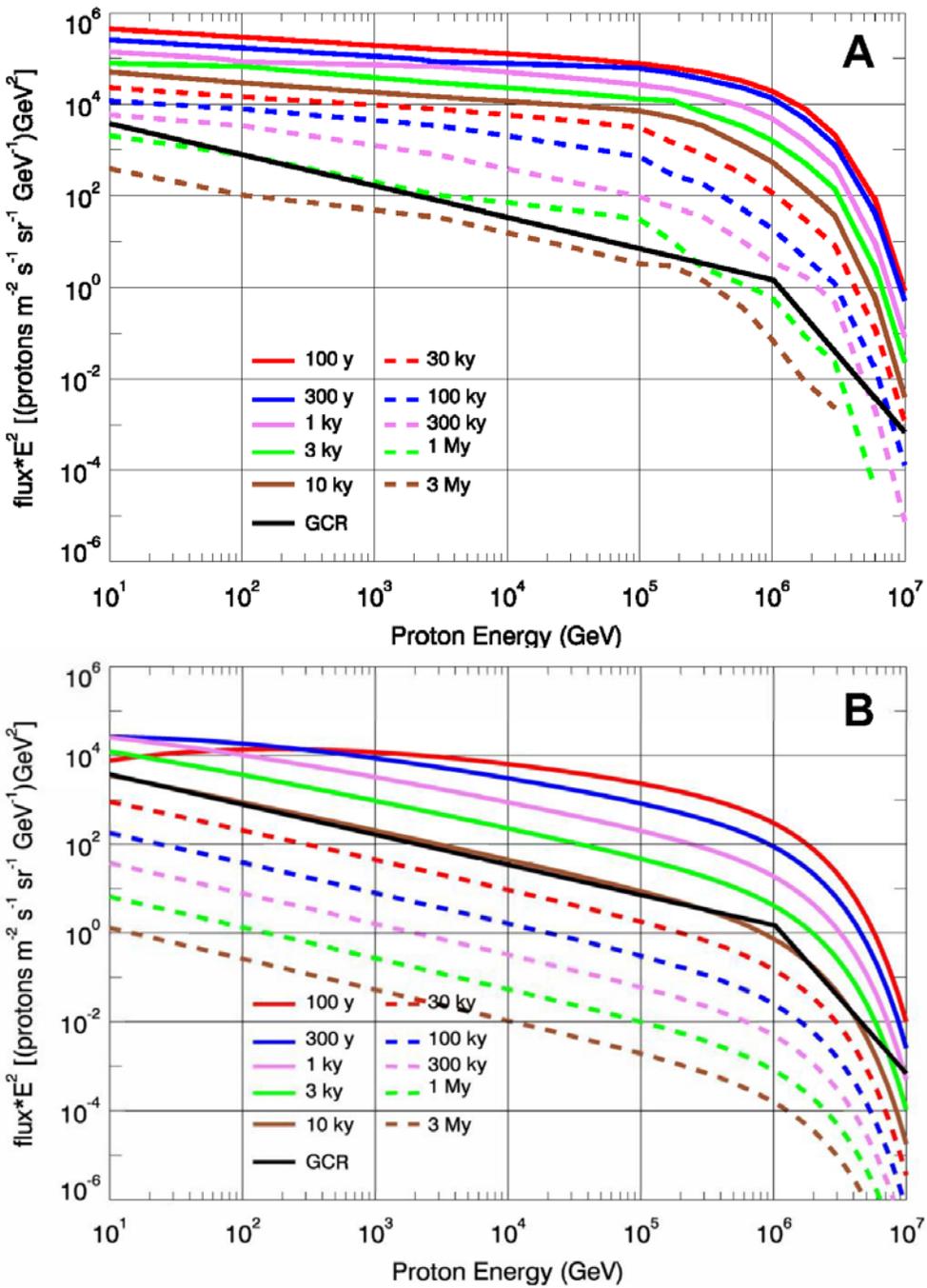

Figure 1

Cosmic ray flux spectrum (times $E^2$) for cases A and B at several times as noted in the figure legend along with Galactic cosmic ray background flux (GCR; black line).

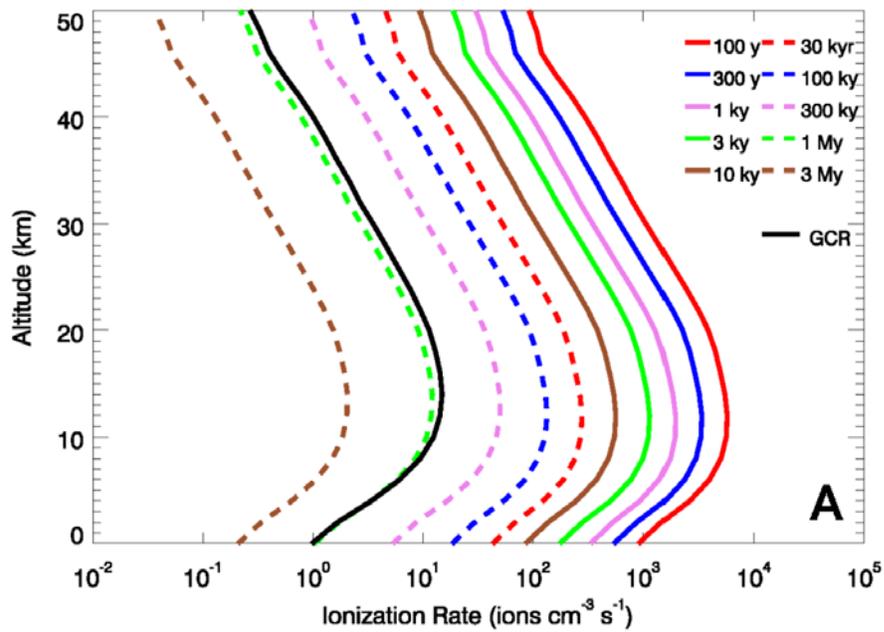

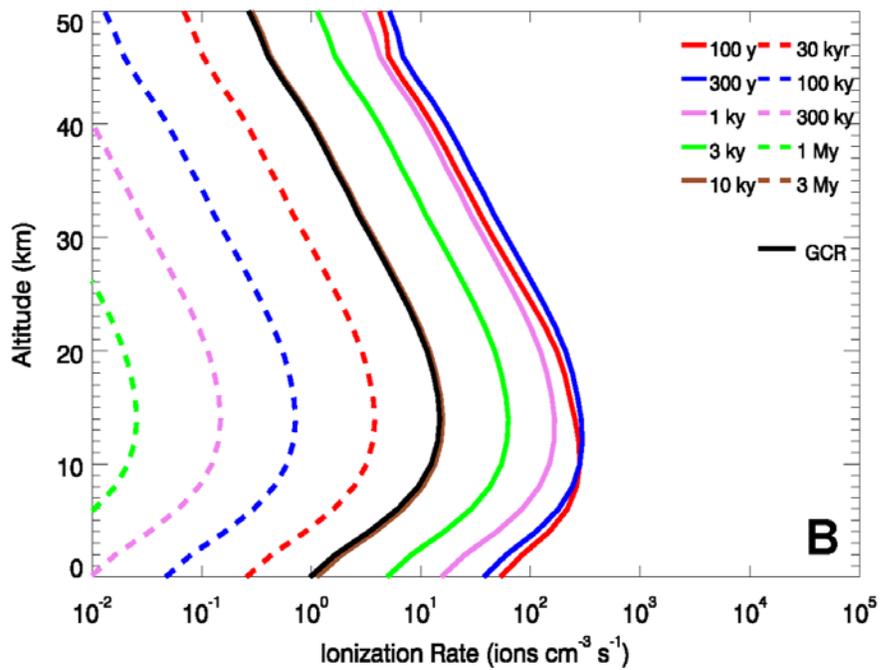

Figure 2

Atmospheric ionization rates for each case at several times, along with Galactic cosmic ray background ionization rate (GCR; black line).

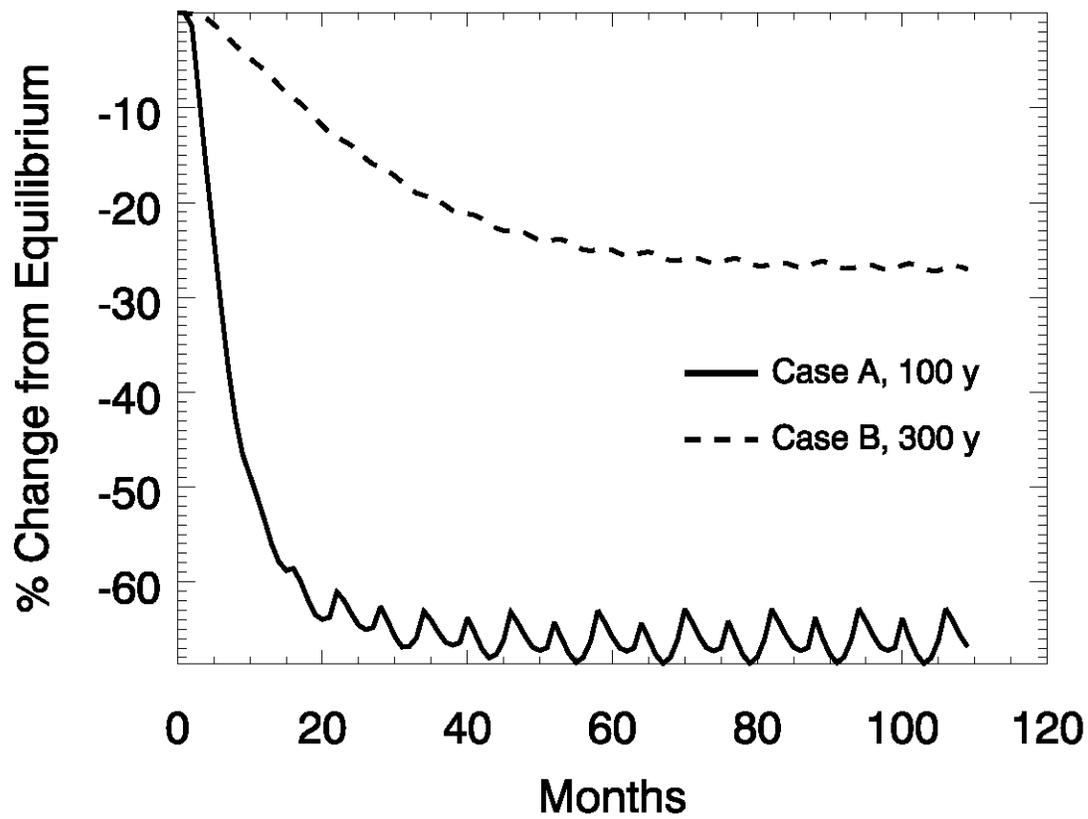

Figure 3

Globally averaged change in atmospheric $O_3$ column density, assuming steady state irradiation in Case A at 100 yrs (solid line) and Case B at 300 yrs (dashed line).

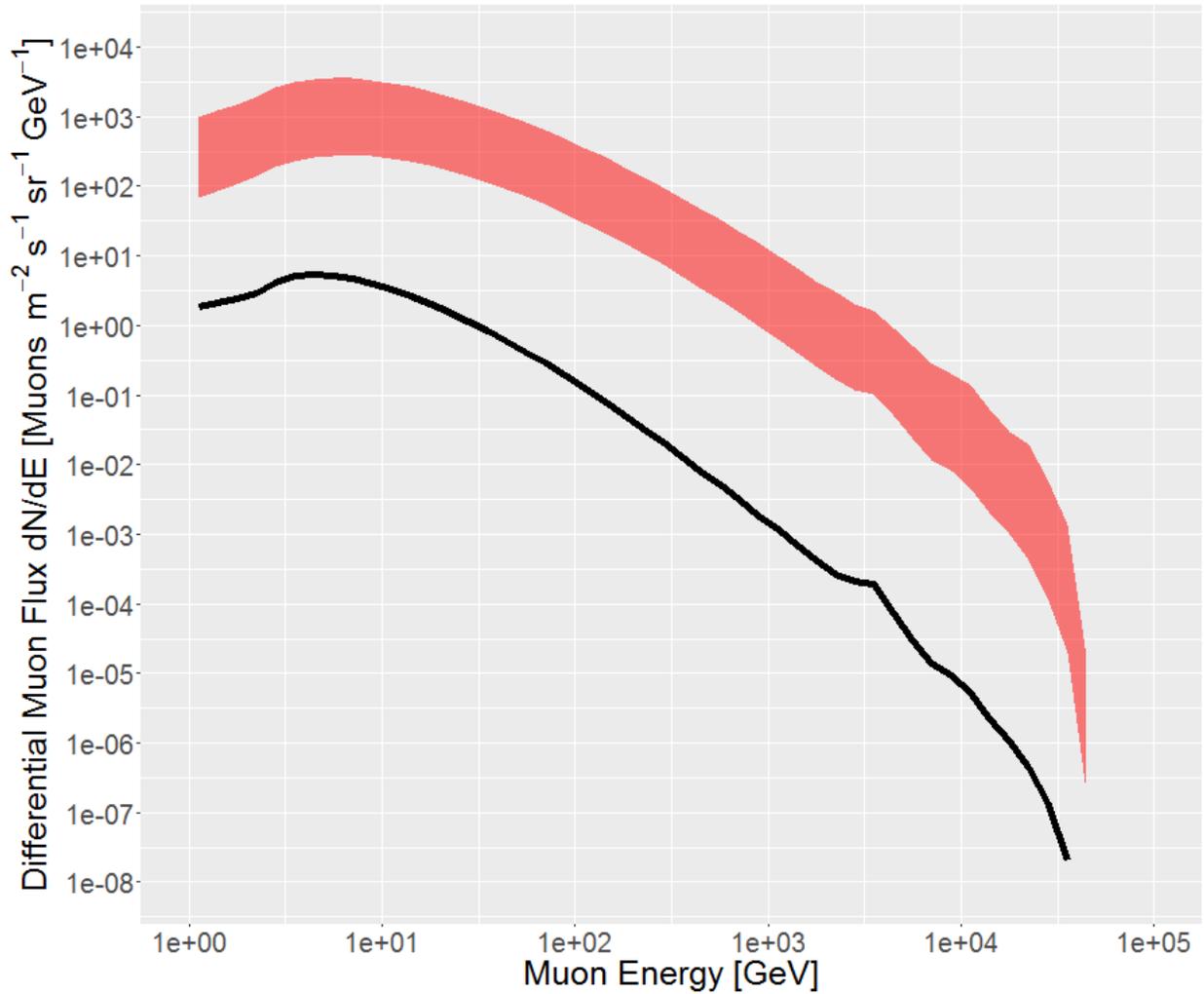

Figure 4

The differential muon spectrum at sea level at time 100 years, when it is at a maximum. The black line represents the typical present spectrum, primarily as a result of showers initiated by galactic cosmic rays. The upper boundary of the banded region represents Case A; the lower boundary represents Case B. Of course, negligible doses are induced in the case of a transverse ordered galactic magnetic field, which is not shown.

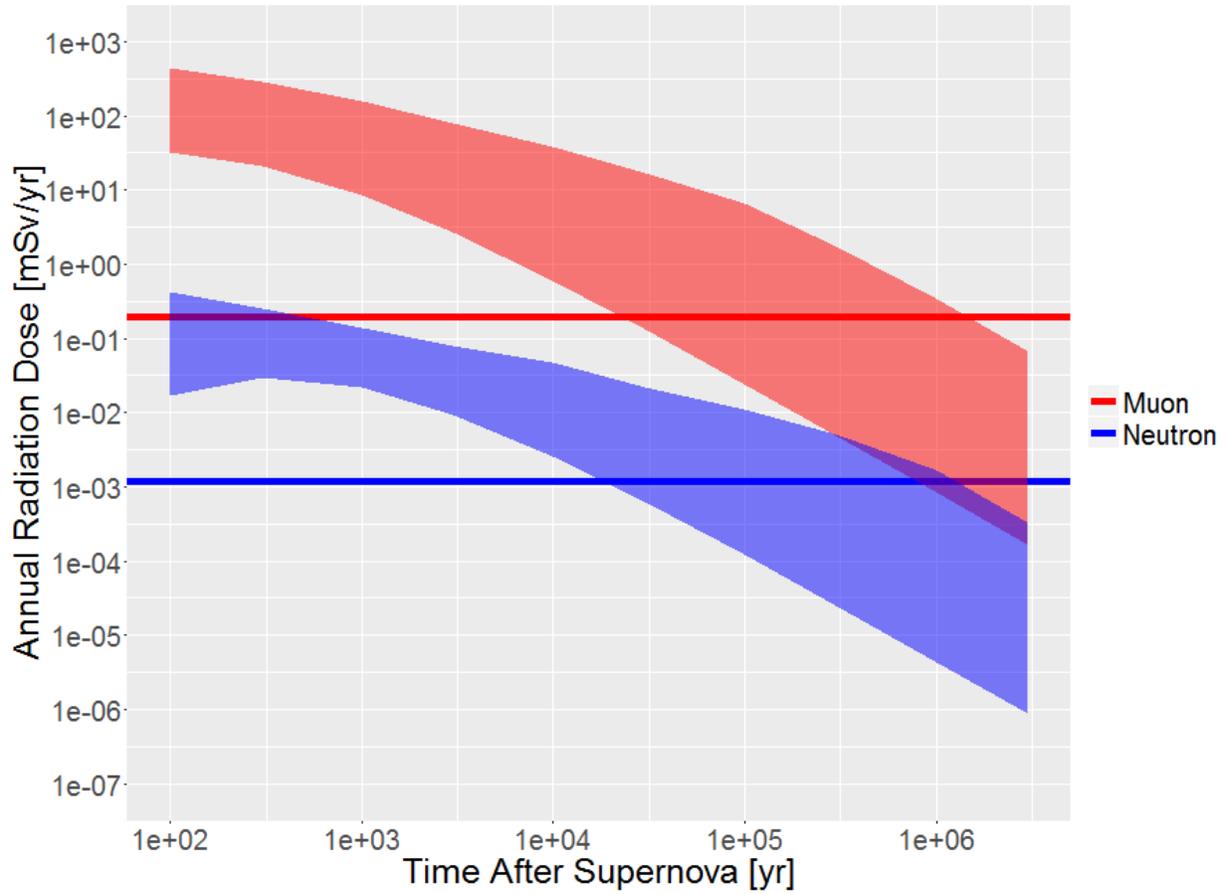

Figure 5

Time evolution of the muon and neutron radiation dose on the ground. The horizontal lines represent the current average dose on the ground. The upper boundary of the banded region represents Case A; the lower boundary represents Case B.